\documentclass[aps,prl,twocolumn,floatfix,superscriptaddress]{revtex4}

\usepackage{times}
\usepackage{graphicx}
\usepackage{epstopdf}

\usepackage{hyperref}
\hypersetup{colorlinks=true,linkcolor=blue,citecolor=blue}

\usepackage{amsmath}
\usepackage{wrapfig}
\usepackage{amssymb}
\usepackage{bbold}
\usepackage{siunitx}
\usepackage{xcolor}
\usepackage{url}

\newcommand{\figref}[1]{\mbox{Fig.~\ref{#1}}}
 
\def\ket#1{\mathinner{|{#1}\rangle}}

\usepackage{mleftright} 

\newcommand{\tabref}[1]{\mbox{Table~\ref{#1}}}

\renewcommand{\eqref}[1]{\mbox{Eq.~(\ref{#1})}}
\newcommand{\Eqref}[1]{\mbox{Equation~(\ref{#1})}}

\newcommand{\ketbra}[2]{\mleft| #1 \rangle \langle #2 \mright|}
\newcommand{\brakket}[3]{\mleft\langle #1\mleft| #2 \mright| #3\mright\rangle}

\newcommand{\sz}{\sigma_z}
\newcommand{\sx}{\sigma_x}

\newcommand{\abs}[1]{\mleft|#1\mright|}

\newcommand{\nn}{\nonumber}

\newcommand{\be}{\begin{equation}}
\newcommand{\ee}{\end{equation}}
\newcommand{\bea}{\begin{eqnarray}}
\newcommand{\eea}{\end{eqnarray}}

\begin{document}

\title{Simulating ultrastrong-coupling processes breaking parity conservation in Jaynes-Cummings systems}

\author{Carlos S\'anchez Mu\~noz}
\affiliation{Theoretical Quantum Physics Laboratory, RIKEN Cluster for Pioneering Research, Wako-shi, Saitama 351-0198, Japan}
\affiliation{Clarendon Laboratory, University of Oxford, Parks Road, Oxford OX1 3PU, United Kingdom}

\author{Anton Frisk Kockum}
\affiliation{Theoretical Quantum Physics Laboratory, RIKEN Cluster for Pioneering Research, Wako-shi, Saitama 351-0198, Japan}
\affiliation{Wallenberg Centre for Quantum Technology, Department of Microtechnology and Nanoscience,
Chalmers University of Technology, 412 96 Gothenburg, Sweden}

\author{Adam Miranowicz}
\affiliation{Theoretical Quantum Physics Laboratory, RIKEN Cluster for Pioneering Research, Wako-shi, Saitama 351-0198, Japan}
\affiliation{Faculty of Physics, Adam Mickiewicz University, 61-614 Pozna{\'n}, Poland}

\author{Franco Nori}
\affiliation{Theoretical Quantum Physics Laboratory, RIKEN Cluster for Pioneering Research, Wako-shi, Saitama 351-0198, Japan}
\affiliation{Department of Physics, The University of Michigan, Ann Arbor, Michigan 48109-1040, USA}

\date{\today}

\begin{abstract}
We propose the effective simulation of light-matter ultrastrong-coupling phenomena with strong-coupling systems. Recent theory and experiments have shown that the quantum Rabi Hamiltonian can be simulated by a Jaynes--Cummings system with the addition of \emph{two} classical drives.
This allows to implement nonlinear processes that do not conserve the total number of excitations. However, parity is still a conserved quantity in the quantum Rabi Hamiltonian, which forbids a wide family of processes involving virtual transitions that break this conservation. Here, we show that these parity-non-conserving processes can  be simulated, and that this can be done in an even simpler setup: a Jaynes-Cummings type system with the addition of a \emph{single} classical drive. By shifting the paradigm from simulating a particular model to simulating a particular process, we are able to implement a much wider family of nonlinear coherent protocols than in previous simulation approaches, doing so with fewer resources and constraints. We focus our analysis on three particular examples: a single atom exciting two photons, frequency conversion, and a single photon exciting two atoms.


\end{abstract}

\maketitle


\emph{Introduction.}---The ultrastrong coupling (USC) of light and matter is attracting increasing interest beyond the fields of cavity~\cite{Haroche2013} and circuit~\cite{Gu2017, Kockum2019a} quantum electrodynamics (QED)~\cite{Kockum2019, Forn-Diaz2019}. This interest has been stimulated in the last decade by several experiments finally reaching USC in a variety of physical systems~\cite{Kockum2019, Forn-Diaz2019}. The USC of light and matter (e.g., a cavity mode and a natural or artificial atom) occurs when their coupling strength $g$ becomes comparable to the atomic ($\omega_a$) or cavity ($\omega_c$) frequencies.  More precisely, according to the usual convention, the USC regime occurs when $\eta = \max( g / \omega_c, g / \omega_a)$ is in the range $[0.1,1)$. The regime $\eta \ge 1$ is often referred to as deep strong coupling (DSC)~\cite{Casanova2010}.

Compared to strong coupling (SC; $\eta < 0.1$, but $g$ larger than the loss rates in the system), USC opens new perspectives for efficiently simulating known effects and observing fundamentally new phenomena in quantum nonlinear optics~\cite{Niemczyk2010, Garziano2015, Ma2015, Garziano2016, Kockum2017, Kockum2017a, Stassi2017, DiStefano2017, DiStefano2018, Macri2018a, Cong2019}, quantum field theory, supersymmetric (SUSY) field theories~\cite{Tomka2015}, cavity optomechanics~\cite{Garziano2015a, Pirkkalainen2015, Benz2016, Macri2016, Cirio2017, Macri2018, DiStefano2019, Settineri2019}, quantum plasmonics~\cite{Tame2013, Benz2016, Todisco2018, Munkhbat2018}, light-induced superconductivity~\cite{Sentef2018, Schlawin2019}, quantum thermodynamics~\cite{Seah2018}, photochemistry (chemistry QED)~\cite{Galego2015, Herrera2016, Ebbesen2016, Martinez-Martinez2018}, as well as metamaterial and material sciences. Ultrastrong coupling also has applications in quantum metrology and spectroscopy~\cite{Ruggenthaler2018}, and quantum information processing~\cite{Nataf2011,Wang2016c,Romero2012,Wang2017,Kyaw2015,Stassi2018,Stassi2017}.

\begin{figure}[t]
\begin{center}
\includegraphics[width=\linewidth]{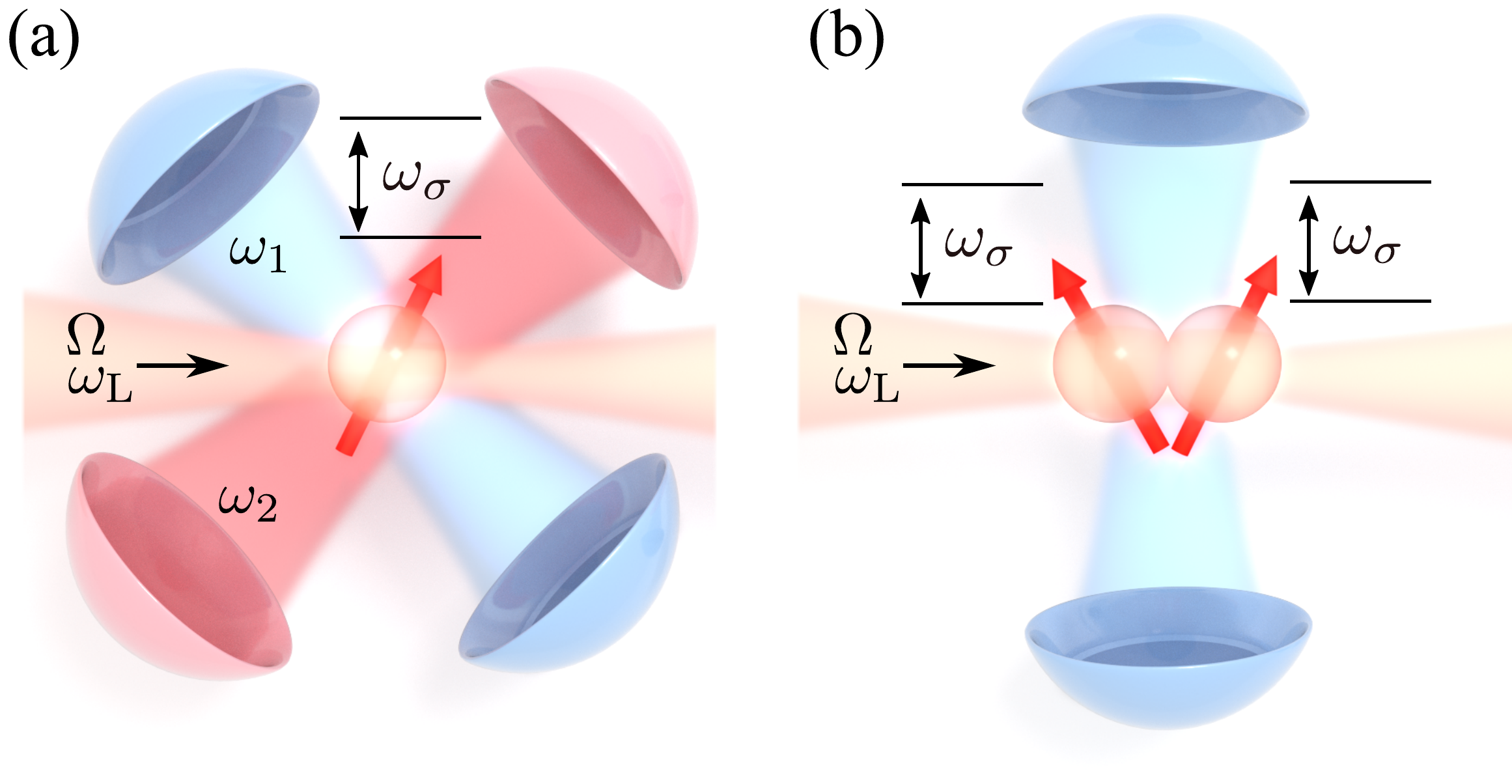}
\end{center}
\caption{Sketches of the two setups that we consider for observing ultrastrong-coupling phenomena. (a) A single two-level atom of frequency $\omega_\sigma$ coupled to two cavities of frequencies $\omega_1$ and $\omega_2$. (b) Two two-level atoms of frequency $\omega_\sigma$ coupled to a cavity of frequency $\omega_a$. In both setups, a single coherent drive of frequency $\omega_\mathrm L$ and amplitude $\Omega$ is applied to each atom.
\label{fig:Setups}}
\end{figure}

The basic model for USC of a single two-level atom to a single-mode cavity is the quantum Rabi model~\cite{Rabi1936, Rabi1937} (QRM). Its multi-atom or multi-mode generalizations include the Dicke~\cite{Dicke1954} and Hopfield~\cite{Hopfield1958} models. When $\eta < 0.1$, these models for USC can be reduced to the simpler Jaynes--Cummings model~\cite{Jaynes1963} (JCM) and its multi-mode or multi-atom generalizations (e.g. the Tavis--Cummings model~\cite{Tavis1968}). Since SC is typically easier to realize in experiment than USC, the question arises whether the predicted USC phenomena can be observed or at least simulated also in the SC regime, e.g., by adding classical drives applied to atom(s) or cavity mode(s) in the SC models. We note that simulating the QRM could also enable simulating other closely related fundamental quantum models, which include the spin-boson~\cite{LeHur2012, Leppakangas2018} and Kondo~\cite{LeHur2012, Goldstein2013, Snyman2015} renormalization-group models, the Rashba-Dresselhaus model~\cite{Tomka2015}, and the Jahn-Teller model~\cite{Hines2004, Meaney2010, Larson2008, Bourassa2009, Dereli2012} among others. Even vacuum-induced symmetry breaking~\cite{Garziano2014}, which is analogous to the Higgs mechanism, has been predicted in the USC regime.  Quantum simulations of the atom-cavity dynamics in the USC and DSC regime in the Rabi and Dicke models have recently attracted much theoretical~\cite{Dimer2007, Ballester2012, Grimsmo2013, Pedernales2015, Felicetti2015, Puebla2017, Fedortchenko2017, Wang2017a, Aedo2018, Qin2018, Leroux2018, Lambert2018, DiPaolo2019} and experimental~\cite{Crespi2012, Braumuller2017, Langford2017, Lv2018} interest. The methods described in Refs.~\cite{Ballester2012, Pedernales2015} and implemented in circuit-QED~\cite{Braumuller2017} and trapped-ion experiments~\cite{Lv2018} simulate the QRM in the USC regime with a light-matter system described by the JCM in the SC regime. These quantum simulations require \emph{two} drives to be applied to a system with a \emph{single atom} and a \emph{single-mode} resonator.

In this Letter, we propose to shift the paradigm from simulating the full QRM to simulating particular processes characteristic of the USC regime, e.g., violating the conservation of number of particles.  This approach allows to employ fewer resources and to go beyond the standard QRM and implement hallmark USC processes that are forbidden due to parity conservation~\cite{Garziano2015, Garziano2016, Kockum2017, Kockum2017a, Stassi2017}. We illustrate this approach by analysing three different phenomena that require breaking parity conservation: a single two-level atom emitting two photons~\cite{Garziano2015}, frequency conversion of two photonic modes coupled to a two-level atom~\cite{Kockum2017a}, and a single photon exciting two atoms~\cite{Garziano2016}.  We also give a protocol for an experimental implementation and show its feasibility in several, well-developed experimental systems.


\emph{Hamiltonians for light-matter coupling.}---The QRM describes the interaction between a two-level atom (qubit) of frequency $\omega_a$ and a cavity mode of frequency $\omega_c$ by the Hamiltonian ($\hbar=1$)
\begin{equation}
H_R = H_0 + \sx X = H_0 + g \mleft( \sigma + \sigma^\dag \mright) \mleft( a + a^\dag \mright),
\label{eq:H_R}
\end{equation}
where $H_0 = (\omega_a / 2) \sz + \omega_c a^\dag a$ is the free Hamiltonian, $a$ ($a^\dag$) is the annihilation (creation) operator of the cavity mode, $X = a + a^\dag$ is the canonical position operator, $\sx = \sigma + \sigma^\dag$ and $\sigma_z$ are Pauli operators, $\sigma$ ($\sigma^\dag$) is the atomic lowering (raising) operator, and $g$ is the atom-field coupling constant. Under the rotating-wave approximation (RWA), which is valid if $\{ \omega_c, \omega_a\} \gg \{ g, \abs{\omega_c - \omega_a} \}$, the counter-rotating terms, $\sigma^\dag a^\dag$ and $\sigma a$, in \eqref{eq:H_R} can be ignored. This leads to the standard JCM described by the Hamiltonian $H_{\rm JC} = H_0 + g \mleft( \sigma a^\dagger + \sigma^\dagger a \mright)$. The counter-rotating terms can be effectively restored in the JCM in various ways, e.g., using cavity-light squeezing~\cite{Qin2018, Leroux2018} to enhance the coupling strength $g$. A simpler method is to apply two time-dependent classical drives, as suggested in Ref.~\cite{Ballester2012}. This yields an effective QRM Hamiltonian in a rotated frame, where the ratio $\eta \equiv g / \omega_c$ can be effectively increased as $\eta' \equiv g' / \omega'_c = g / [2 (\omega_c - \omega_1) ]$  (with $\omega_1$ one of the driving frequencies) from the SC regime up to the USC regime, or even the DSC regime. In this scheme, the effective frequency of the qubit is equal to the amplitude of one of the drives.

\Eqref{eq:H_R}, however, does not include any term that changes the number of particles by an odd number, and thus parity is conserved. We now show how a simpler JCM setup, with only a \emph{single} drive, can be used to simulate any particular process characteristic of the USC, i.e., relying on counter-rotating terms in the QRM, but also violating parity conservation through terms of the type $\sigma_z(a+a^\dagger)$. Our approach is inspired by earlier work on creating multi-photon states in cavity QED~\cite{SanchezMunoz2014, SanchezMunoz2015,SanchezMunoz2018}.


\emph{USC process I: Two photons excited by a single atom.}---We first consider the setup in \figref{fig:Setups}(a), i.e., two cavities coupled to a single qubit that is coherently driven by a classical field. In a frame rotating with the frequency $\omega_\mathrm L$ of the driving field, the Hamiltonian is given by
\begin{eqnarray}
H &=& \Delta_1 a_1^\dag a_1 + \Delta_2 a_2^\dag a_2 + \Delta_\sigma \sigma^\dag \sigma + \Omega \mleft( \sigma + \sigma^\dag \mright) \nn\\
&& + g \mleft[ \sigma \mleft( a_1^\dag + a_2^\dag \mright) + \mathrm{h.c.} \mright],
\label{eq:hamiltonian_two_photons}
\end{eqnarray}
with $a_{1,2}$ the bosonic annihilation operators of the cavity modes; $\Delta_1$, $\Delta_2$ and $\Delta_\sigma$ are the frequency detunings between the cavities or qubit and the drive ($\Delta_x \equiv \omega_x - \omega_\mathrm L$), $\Omega$ is the amplitude of the driving field, and $g$ is the coupling rate between the cavities and the qubit (considered to be equal for simplicity). 

The part of \eqref{eq:hamiltonian_two_photons} that only depends on $\sigma$ can be easily diagonalized. Denoting the ground and excited eigenstates of an undriven qubit $\ket{g}$ and $\ket{e}$, respectively, the new eigenstates with the driving applied correspond to a rotated spin basis, i.e.
\begin{eqnarray}
\ket{+} &=& \cos \theta \ket{g} + \sin \theta \ket{e} = e^{i \sigma_y 2 \theta} \ket{g}, \\
\ket{-} &=& \sin \theta \ket{g} - \cos \theta \ket{e} =  - e^{i \sigma_y (2 \theta + \pi)} \ket{g},
\end{eqnarray}
with $\cos \theta \equiv 1 / \sqrt{1 + \xi^{-2}}$, $\sin \theta \equiv 1 / \sqrt{1 + \xi^2}$, $\theta \in [0, \pi/2]$, and $\xi \equiv \Omega / (\Delta_\sigma / 2 + R)$, where $R$ is the Rabi frequency given by $R \equiv \sqrt{\Omega^2 + (\Delta_\sigma / 2)^2}$. 

Working in the eigenbasis $\ket{\pm}$, the original lowering operator $\sigma$ can be written in terms of the new operators $\tilde \sigma \equiv \ketbra{-}{+}$ as $\sigma = s^2 \tilde \sigma - c^2 \tilde \sigma^\dag + c s \tilde \sigma_z$, with $s = \sin \theta$, $c = \cos \theta$, $\tilde \sigma_{z} \equiv 2 \tilde \sigma^\dag \tilde \sigma - \mathbb{1}$.
Therefore, the resulting Hamiltonian in the rotated spin basis reads
\begin{eqnarray}
H &=& \Delta_1 a_1^\dag a_1+\Delta_2 a_2^\dag a_2 + R \tilde \sigma_z \nn\\
&& + g \mleft[ \mleft( s^2 \tilde \sigma - c^2 \tilde \sigma^\dag + c s \tilde \sigma_z \mright) \mleft( a_1^\dag + a_2^\dag \mright) + \mathrm{h.c.} \mright]\,.
\label{eq:H-rotated}
\end{eqnarray}
The transition energy of the effective qubit is now given by $R$, which can be made small enough that counter-rotating terms of the kind $\tilde \sigma^\dag a_1^\dag$ and $\tilde \sigma_z a_1^\dag$ play a relevant role in the dynamics. The presence of the latter type of coupling terms, involving $\tilde\sigma_z$, makes $H$ reminiscent of the generalized QRM, where a coupling term proportional to $\sz \mleft( a + a^\dag \mright)$ is added to the QRM in \eqref{eq:H_R}~\cite{Garziano2015, Garziano2016, Kockum2017, Deppe2008, Niemczyk2010, Yoshihara2017}.  Crucially, the presence of the $\tilde\sigma_z$ coupling term, which was absent in previous proposals of simulation of the QRM, breaks parity symmetry and enables processes that changes the number of excitations in the system by an odd number~\cite{Kockum2017, Kockum2019, Forn-Diaz2019}.

We will now see how, in the limit of ${\alpha \equiv g / R \ll 1}$, the counter-rotating terms in \eqref{eq:H-rotated} lead to Rabi oscillations between pairs of eigenstates of the bare Hamiltonian that are \emph{not} directly coupled by the interactions~\cite{Kockum2017}, with Rabi frequencies $\propto \alpha g$. In the effective USC regime when $\alpha \sim 0.1$, we find  the optimal condition in which $g / R \ll 1$ remains valid, while the effective Rabi frequencies $\sim 0.1 g$ can be significant compared to decoherence rates. The normal USC condition $\eta \gtrsim 0.1$ for observing these phenomena is thus lifted.

\begin{figure*}
\begin{center}
\includegraphics[width=\linewidth]{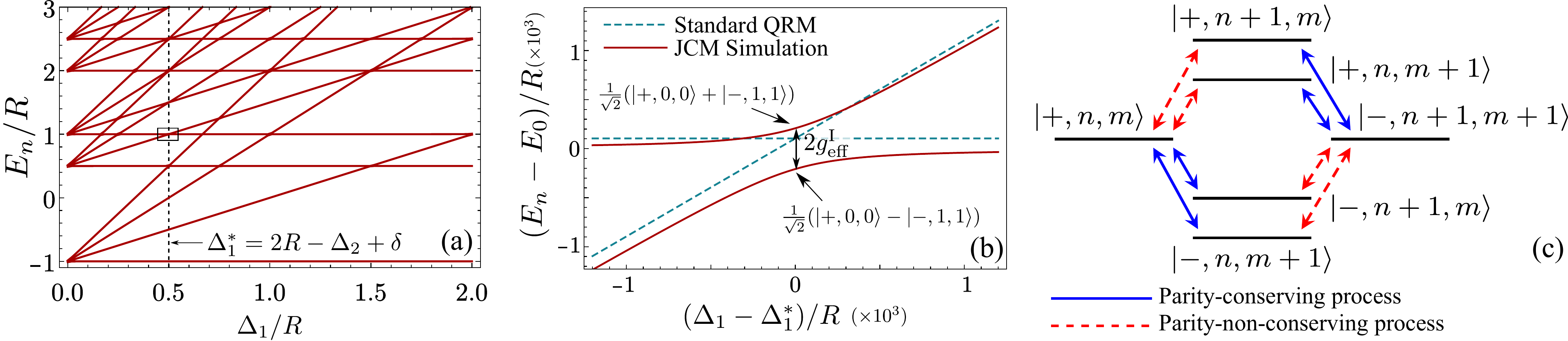}
\end{center}
\caption{USC process I: Energy-level diagrams and transitions for the process where a single atom emits two photons.
(a) Energy levels $E_n$ for the Hamiltonian in \eqref{eq:H-rotated} as a function of $\Delta_1$. Parameters: $\Omega = 80g$, $\Delta_\sigma = \Omega/\sqrt{2}$,  $\Delta_2 = 3R/2$, with $R=\sqrt{\Omega^2+\Delta_\sigma^2/4}$. The Hilbert space is truncated at three photons for simplicity.
(b) Zoom-in on the anti-crossing between the energy levels corresponding to $\ket{+,0, 0}$ and $\ket{-,1, 1}$. The size of the level splitting at the resonance $\Delta_1 = \Delta_1^* \approx 2R-\Delta_2$---where both states have the same energy in the absence of coupling  $E_0 =\Delta_2-R= R$---indicates the strength $g_\mathrm{eff}^\mathrm{I}$ of the effective interaction between these two states.
(c) The transitions in the second-order process that creates the effective coupling between $\ket{+, n, m}$ and $\ket{-, n+1, m+1}$. Red dashed (blue solid) arrows indicate transitions that change the total number of excitations in the system by one (zero), i.e., transitions mediated by counter-rotating (non-rotating) terms. Crucially, the whole process cannot occur without involving intermediate, parity-non-conserving processes.
\label{fig:eigenvalues}}
\end{figure*}

One example of such a nonlinear process is the simultaneous excitation of one photon in each cavity by the single qubit. By plotting the energy levels of \eqref{eq:H-rotated} [\figref{fig:eigenvalues}(a)], and zooming in around $\Delta_1 + \Delta_2 \approx 2R$, we find an avoided-level crossing [\figref{fig:eigenvalues}(b)]. The interaction around this point is described by the effective Hamiltonian~\cite{SupMat}:
\begin{eqnarray}
&& H_\mathrm{eff}^\mathrm{I} = \Delta_1 a_1^\dag a_1 + \Delta_2 a_2^\dag a_2 + \mleft( R + \lambda \mright) \tilde \sigma_z \nn\\
&& + \mleft( \chi_1 a^\dag_1 a_1 + \chi_2 a^\dag_2 a_2 \mright) \tilde \sigma_z + g_\mathrm{eff}^\mathrm{I} \mleft( a_1^\dag a_2^\dag \tilde \sigma + \mathrm{h.c.} \mright) \,, \quad
\label{eq:Heff1}
\end{eqnarray}
which couples the states $\ket{+, n, m} \leftrightarrow \ket{-, n+1, m+1}$, confining the dynamics inside that manifold. This effective interaction requires both states to be quasi-resonant, which implies, ignoring for now small dispersive energy shifts, the two conditions:
\begin{equation}
\Delta_1 + \Delta_2 \approx 2R, \quad  \Delta_1 \neq \Delta_2 \neq \mleft( \pm R, \pm 2R \mright).
\label{eq:resonance-condition1}
\end{equation}
The second condition is imposed in order to be detuned from first-order processes (e.g., $\tilde \sigma a_1^\dag + \text{h.c.}$ if $\Delta_1 = 2 R$) and competing second-order processes [e.g., $\left(\tilde \sigma {a_1^\dag}^2 + \text{h.c.}\right)$ for $\Delta_1 = R$] exciting degenerate photon pairs within a single cavity~\cite{SanchezMunoz2014,SanchezMunoz2015, SanchezMunoz2018}. The effective two-photon coupling rate in \eqref{eq:Heff1} is given by
\begin{equation}
g_\mathrm{eff}^\mathrm{I} = g^2 c s^3[R f (1 - f)]^{-1},
\label{eq:geffI}
\end{equation}
where we defined $\Delta_1 = 2 f R$ and $\Delta_2 = (1 - f) 2 R$, $f \in (0,1)$, so that \eqref{eq:resonance-condition1} is automatically fulfilled. This effective interaction is mediated by the second-order processes shown in \figref{fig:eigenvalues}(c). The Lamb shift of the qubit is

\begin{equation}
\lambda = g^2 \mleft[c^4 \mleft( \Delta^{-1}_{1,+}+\Delta^{-1}_{2,+} \mright) - s^4 \mleft(\Delta^{-1}_{1,-}+ \Delta^{-1}_{2,-} \mright) \mright]/2,
\end{equation}
and the dispersive coupling rates are
\begin{equation}
\chi_i = g^2 \mleft( c^4/\Delta_{i,+}- s^4/\Delta_{i,-}\mright),
\end{equation}
with $\Delta_{i,\pm}\equiv \Delta_i \pm 2 R$.

\Eqref{eq:geffI} shows that the resonant-driving condition $\Delta_\sigma = 0$ ($\theta = \pi / 4$) does not provide the maximum possible two-photon coupling rate. In particular, for a fixed $R$, we see that the optimal angle $\theta$ maximizing $g_\mathrm{eff}^\mathrm{I}$ is $\theta^* = \pi / 3$. This angle yields the optimum value $g_\mathrm{eff}^\mathrm{I} (\theta^*)\approx 1.3 g_\mathrm{eff}^\mathrm{I} (\theta = \pi / 4)$.
Alternatively, we can compute the optimal detuning $\Delta_\sigma$ for a fixed $\Omega$, which is experimentally more meaningful since varying $\Delta_\sigma$ for a fixed $\Omega$ is more straightforward than varying $\theta$ for a fixed $R$. By writing \eqref{eq:geffI} explicitly in terms of $\Delta_\sigma$ and $\Omega$, we obtain the optimal detuning $\Delta_\sigma^* = \Omega / \sqrt{2}$.
The corresponding value of $g_\mathrm{eff}^\mathrm{I}$ is then given by $g_\mathrm{eff}^\mathrm{I} (\Delta_\sigma^*) \approx 1.18 g_\mathrm{eff}^{\mathrm{I}} (\Delta_\sigma = 0)$.

In order to obtain full two-photon Rabi oscillations between the  states $\ket{1} = \ket{+, n, m}$ and $ \ket{2} = \ket{-, n+1, m+1}$, the quasi-resonance condition \eqref{eq:resonance-condition1} needs to be fine-tuned to account for the Lamb shift of the qubit and the dispersive qubit-cavity couplings in \eqref{eq:Heff1}, given by $\lambda$ and $\chi_1, \chi_2$. In other words, $\Delta_1$ and $\Delta_2$ must be chosen such that $\brakket{1}{H_\mathrm{eff}}{1} = \brakket{2}{H_\mathrm{eff}}{2}$. Introducing a correction $\delta$ such that $\Delta_1 = 2 R f + \delta$, we solve this equation for the Hamiltonian in \eqref{eq:Heff1} and obtain
\begin{equation}
\delta = 2 \lambda + \chi_1 \mleft( 2 n + 1 \mright) + \chi_2 \mleft( 2 m + 1 \mright).
\label{eq:delta-shift1}
\end{equation}
%

\begin{figure}[t]
\begin{center}
\includegraphics[width=\linewidth]{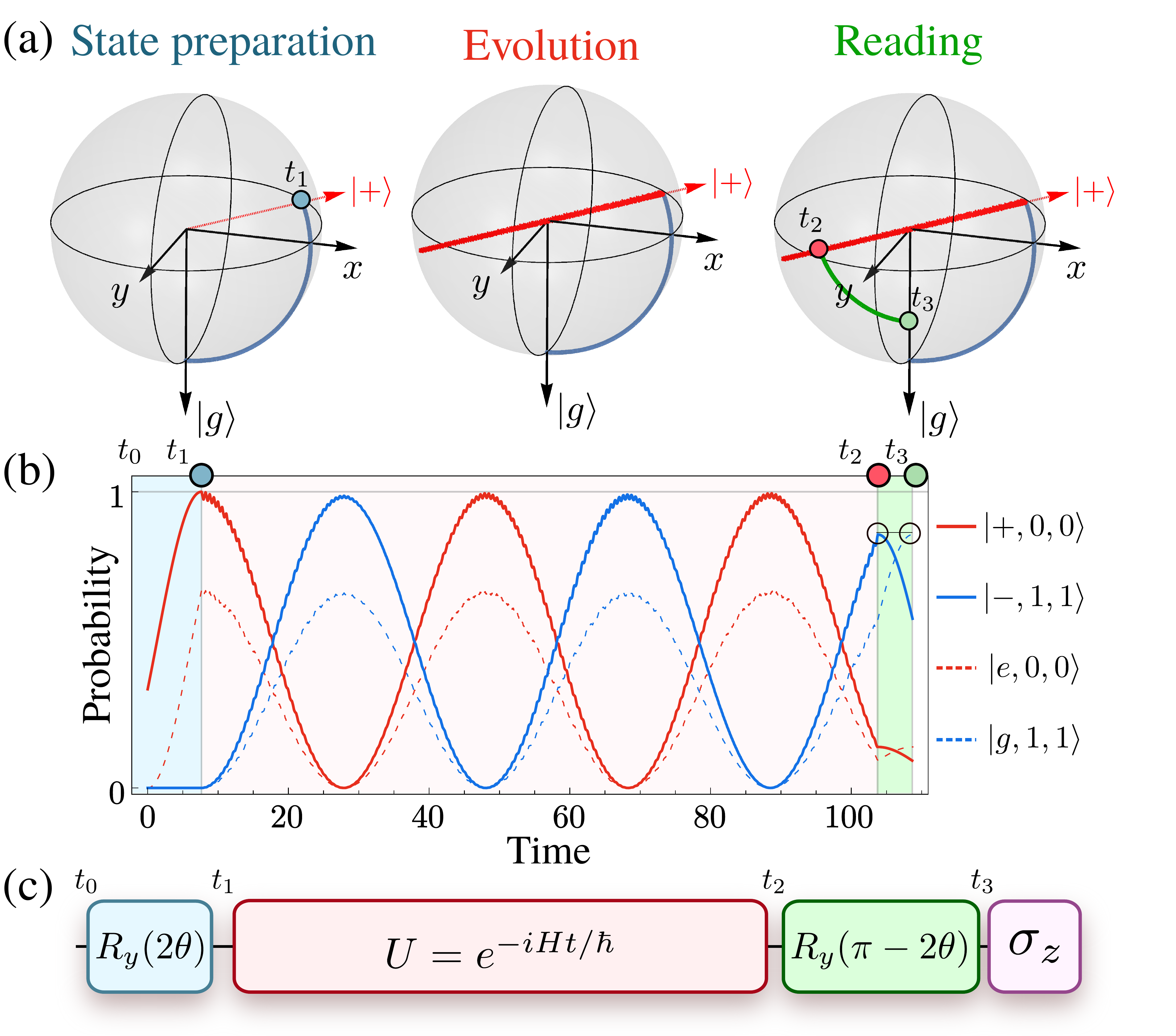}
\end{center}
\caption{USC process I: Illustration of the experimental protocol, showing (a) the Bloch sphere picture of the qubit state, (b) the time evolution of the occupation probabilities for the most relevant states, and (c) a schematic diagram of the proposed pulse sequence. From time $t_0$ to $t_1$, the qubit is rotated to the correct initial state in the rotated spin basis. From time $t_1$ to $t_2$, the qubit is driven and the system evolves according to the Hamiltonian in \eqref{eq:H-rotated}, i.e., moving back and forth along the red trajectory depicted in the Bloch sphere. At time $t_2$, the drive is turned off and the qubit is rotated back to the original basis, where it is then measured at time $t_3$.
\label{fig:protocol}}
\end{figure}
\emph{Experimental protocol.}---We now discuss an experimental protocol for implementing and measuring the non-linear process. This protocol is shown in \figref{fig:protocol}. Starting with no photons in the cavities and the qubit in its ground state $\ket{g}$, the first step is to apply a rotation of $2 \theta$ around the $y$-axis to bring the qubit into the eigenstate $\ket{+}$. This single-qubit rotation is a basic element of any quantum information toolbox. At this stage, the cavities and the qubit are detuned and no interaction takes place. Then, the driving field is switched on and the nonlinear process becomes resonant. After the system has evolved for a time $t$, the drive is switched off (effectively decoupling the qubit and the cavity), and the state of the qubit in the $\ket{\pm}$ basis is transformed back into the $\mleft \{ \ket{g}, \ket{e} \mright\}$ basis (eigenstates of $\sz$) by applying a rotation of $(\pi - 2 \theta)$ around the $y$-axis. A measurement of the qubit population in the $\mleft \{ \ket{g}, \ket{e} \mright\}$ basis then reveals the qubit final state in the rotated basis.


\emph{USC process II: Frequency conversion.}---The setup in \figref{fig:Setups}(a) can also be exploited to engineer other processes, e.g., frequency conversion. In that case, we want to couple the states $\ket{n+1, m, -}$ and $ \ket{n, m+1, +}$. The resonance conditions then become
\begin{equation}
\Delta_1 \approx  2 R + \Delta_2, \quad  \Delta_1 \neq \Delta_2 \neq \pm R
\end{equation}
where, again, the second condition guarantees that second-order processes introducing photon pairs into the cavities are off-resonance. Following the same procedure outlined in~\cite{SupMat}, we obtain the effective Hamiltonian
\begin{eqnarray}
&& H_\mathrm{eff}^\mathrm{II} = \Delta_1 a_1^\dag a_1 + \Delta_2 a_2^\dag a_2 + (R + \lambda) \tilde \sigma_z \nn\\
&& + \mleft( \chi_1 a^\dag_1 a_1 + \chi_2 a^\dag_2 a_2 \mright) \tilde \sigma_z + g_\mathrm{eff}^\mathrm{II} \mleft( a_1^\dag a_2 \tilde \sigma + \mathrm{h.c.} \mright) \, ,
\label{eq:Heff2}
\end{eqnarray}
where the frequency-conversion rate is given by
\begin{equation}
g_\mathrm{eff}^\mathrm{II} =g^2 \mleft[ (f - 1) c^3 s + f c s^3 \mright][R f (f - 1)]^{-1} ,
\end{equation}
having now defined $\Delta_1 = 2 f R$ and $\Delta_2 = (f-1) 2 R$, $f \in (0,1)$.
Once again, driving the qubit on resonance does not maximize $g_\mathrm{eff}^\mathrm{II}$. Frequency-conversion-rate increases by 50-70\% compared to resonant driving can be achieved by using the optimal angle $\theta^*$ or the optimal detuning $\Delta^*$, whose analytical expressions can be found in the Supplemental Material~\cite{SupMat}.

\emph{USC process III: Two atoms excited by a single photon.}---The last process that we demonstrate is the excitation of two atoms by a single photon, i.e. the direct coupling between the states $\ket{+, +, n}$ and $\ket{-, -, n+1}$. We now consider the setup in \figref{fig:Setups}(b), i.e. a cavity coupled to two coherently driven qubits, with lowering operators $\sigma_{1,2}$. For simplicity and without loss of generality, we consider both qubits to have the same transition frequencies. In the rotating frame of the driving, the Hamiltonian is
\begin{eqnarray}
H &=& \Delta_a a^\dag a + \Delta_\sigma \mleft( \sigma_1^\dag \sigma_1 + \sigma_2^\dag \sigma_2 \mright) + \Omega \mleft( \sigma_1 + \sigma_2 + \mathrm{h.c.} \mright) \nn\\
&& + g \mleft[ a \mleft( \sigma_1^\dag + \sigma_2^\dag \mright) + \mathrm{h.c.} \mright] \, ,
\label{eq:initial-H}
\end{eqnarray}
where $a$ is the bosonic annihilation operator of the cavity, and $\Delta_a=\omega_a-\omega_\mathrm L$ ($\Delta_\sigma=\omega_\sigma-\omega_\mathrm L$) is the cavity (qubit) detuning from the drive frequency. In the dressed-qubits basis, the resonance condition enabling the desired non-linear process simply reads $\Delta_a \approx 4 R$. We then obtain~\cite{SupMat} the effective Hamiltonian
\begin{eqnarray}
H_\mathrm{eff}^\mathrm{III} &=& \Delta_a a^\dag a + \sum_i (R + \lambda) \tilde \sigma_{z, i} + \chi a^\dag a \tilde \sigma_{z, i} \nn\\
&& + g_\mathrm{eff}^\mathrm{III} \mleft( a^\dag \sigma_1 \sigma_2 + a \sigma_1^\dag \sigma_2^\dag \mright) , 
\end{eqnarray}
with an effective coupling rate that emerges from third-order processes,
\begin{equation}
g_\mathrm{eff}^\mathrm{III} = g^3 \mleft( c^3 s^3 + 3 c s^5 \mright)/(3 R^2).
\label{eq:geff}
\end{equation}
As in the previous cases, the effective coupling can be maximized by driving the qubit slightly off resonance, using either the optimal angle $\theta^*$ or the optimal detuning $\Delta^*$, whose expressions we provide in~\cite{SupMat}.

\begin{table}
\renewcommand{\arraystretch}{1.15}
\renewcommand{\tabcolsep}{0.1cm}
\begin{tabular}{l | c | c | c }
System & $g / (2 \pi)$ & $\gamma / (2 \pi)$ & $(g_\mathrm{eff}^{\mathrm{I}}, g_\mathrm{eff}^{\mathrm{II}}, g_\mathrm{eff}^{\mathrm{III}}) / \gamma$ \\ 
\hline 
Natural atoms~\cite{Birnbaum2005} & \SI{34}{\mega\hertz} &\SI{4.1}{\mega\hertz} & (0.6,\,0.4, \,0.004) \\ 
Trapped ions~\cite{Haffner2008} & \SI{10}{\kilo\hertz}   & \SI{100}{\hertz} & (7.9,\, 5.1, \, 0.05) \\ 
Quantum acoustics~\cite{Arrangoiz-Arriola2019} & \SI{16}{\mega\hertz} & \SI{0.6}{\mega\hertz}   & (2.1,\, 1.3,\, 0.01) \\
Circuit QED~\cite{Magnard2018} & \SI{335}{\mega\hertz} & \SI{0.5}{\mega\hertz}&(52.9,\,33.8,\,0.36) \\ 
Quantum dots~\cite{Arakawa2012} & \SI{19.3}{\giga\hertz} & \SI{6.0}{\giga\hertz} & (0.3,\, 0.2,\,0.002)
\end{tabular}
\caption{Experimentally feasible effective rates for the three processes discussed in the text: (I) a single photon exciting two atoms, (II) frequency conversion, and (III) a single atom exciting two photons. We set $\Omega / g = 20$; $\gamma$ refers to the largest decoherence rate in the system.}
\label{table:experiments}
\end{table}

\emph{Experimental implementations.}---The results presented here are based on very fundamental models that describe the exchange of single excitations between a qubit and a harmonic oscillator, and can, therefore, be applied in many different systems. In Table~\ref{table:experiments}, we compare, under experimentally feasible assumptions, the effective coupling strengths $g_{\rm eff}$ and decoherence rates $\gamma$ that can be obtained in five experimental platforms. An experimental implementation is feasible when $g_{\rm eff} / \gamma > 1$, i.e. when the effective coupling is strong. Table~\ref{table:experiments} shows that the second-order processes that we have proposed here should be ready for implementation in several systems, and the third-order process may be within reach for circuit-QED setups. However, we note that the nonlinear processes may also be exploited even in the dissipative regime where $g_\mathrm{eff} < \gamma$, e.g. yielding multi-photon emission with non-classical properties~\cite{SanchezMunoz2014,SanchezMunoz2015,SanchezMunoz2018,Chang2016}.


\emph{Conclusions.}---We have shown how, rather than simulating the full QRM with a JCM, one can simulate specific nonlinear processes characteristic of the USC regime. This requires fewer resources and allows to implement processes that are forbidden in the standard QRM due to parity conservation. Our method is ready for its implementation on several existing experimental platforms and opens up new possibilities for exploring USC physics and its applications in technologies such as quantum information and quantum metrology.



\paragraph*{Acknowledgements.}

C.S.M. is funded by the Marie Sklodowska-Curie Fellowship QUSON (Project  No. 752180). 
A.F.K. acknowledges support from the Japan Society for the Promotion of Science (BRIDGE Fellowship BR190501), the Swedish Research Council (grant number 2019-03696), and the Knut and Alice Wallenberg Foundation.
A.M. was supported by the Polish National Science Centre
(NCN) under the Maestro Grant No. DEC-2019/34/A/ST2/00081.
F.N. is supported in part by the:
MURI Center for Dynamic Magneto-Optics via the
Air Force Office of Scientific Research (AFOSR) (FA9550-14-1-0040),
Army Research Office (ARO) (Grant No. Grant No. W911NF-18-1-0358),
Asian Office of Aerospace Research and Development (AOARD) (Grant No. FA2386-18-1-4045),
Japan Science and Technology Agency (JST) (via the Q-LEAP program, and the CREST Grant No. JPMJCR1676),
Japan Society for the Promotion of Science (JSPS) (JSPS-RFBR Grant No. 17-52-50023, and
JSPS-FWO Grant No. VS.059.18N), the RIKEN-AIST Challenge Research Fund,
the Foundational Questions Institute (FQXi), and the NTT PHI Laboratory.



\let\oldaddcontentsline\addcontentsline
\renewcommand{\addcontentsline}[3]{}
\bibliographystyle{mybibstyle}
\bibliography{References}
\let\addcontentsline\oldaddcontentsline

\clearpage
\onecolumngrid
\appendix

\begin{center}
{\bf \large Supplementary Material}
\end{center}

\renewcommand{\theequation}{S\arabic{equation}}

\renewcommand{\thefigure}{S\arabic{figure}} 
\setcounter{figure}{0} 
\setcounter{equation}{0}   

\tableofcontents

\section{Effective Hamiltonians}

In this work, we use a matrix form of perturbation theory that allows one to obtain energy corrections to arbitrary orders with a single matrix inversion. Let us consider a Hilbert subspace $\mathcal A$ consisting of $\mathcal N_A$ states $\{|a_1\rangle,|a_2\rangle,\ldots \}$ whose effective dynamics we wish to describe. This subspace is coupled to another subspace $\mathcal B$ consisting of $\mathcal{N}_B$ states $\{|b_1\rangle,|b_2\rangle,\ldots \}$ that we want to adiabatically eliminate. We define the projectors onto the respective subspaces as $P_{\mathcal{A}}$ and $P_{\mathcal{B}}$. The total Hamiltonian of the combined system is given by 
\begin{equation}
H =  \begin{pmatrix} h & V \\ V^\dagger & \tilde H \ \end{pmatrix},
\label{eq:hamiltonian-initial}
\end{equation}
where $h\equiv  P_{\mathcal{A}} H P_{\mathcal{A}}$ is an $(N_\mathcal{A}\times N_\mathcal{A} )$ matrix acting only on $\mathcal A$, $\tilde H \equiv P_{\mathcal{B}} H P_{\mathcal{B}}$ is an $(N_\mathcal{B}\times N_\mathcal{B})$ matrix acting only on $\mathcal B$,  and $V\equiv P_{\mathcal{B}} H P_{\mathcal{A}}$ is an $(N_\mathcal{A}\times N_\mathcal{B})$ matrix coupling both subspaces. Our objective is to obtain an effective Hamiltonian $h^\mathrm{eff}$ describing the dynamics within $\mathcal A$. The underlying assumption is that the eigenvalues of $h$ are close to the energy $E$, while the eigenvalues of $\tilde H$ are detuned from $E$ by values much larger than the elements of $V$, and therefore can be adiabatically eliminated. This is done by writing the eigenvalue problem:
\begin{equation}
\begin{pmatrix} h & V \\ V^\dagger & \tilde H \ \end{pmatrix} \begin{pmatrix} \phi \\ \chi \end{pmatrix} = E \begin{pmatrix} \phi \\ \chi \end{pmatrix},
\end{equation}
where $\phi$ and $\chi$ are column vectors of length $N_\mathcal{A}$ and $N_\mathcal{B}$, respectively. After matrix multiplication, we obtain the following system of two equations for $\phi$ and $\chi$:
\begin{subequations}
\begin{eqnarray}
\label{eq:system1}
(E-h)\phi &=& V \chi,\\
\label{eq:system2}
(E-\tilde H)\chi &=& V^\dagger \phi.
\end{eqnarray}
\end{subequations}
By solving \eqref{eq:system2} and substituting into \eqref{eq:system1}, we obtain:
\begin{equation}
(E-H_\mathrm{eff})\phi = 0,
\end{equation}
where $H_\mathrm{eff}(E) = h + \delta h$, and
\begin{equation}
\delta h = V\frac{1}{E-H}V^\dagger.
\label{eq:effective-H}
\end{equation}
$H_\mathrm{eff}$ corresponds to effective Hamiltonians that we have presented in the main text. Notably, this simple expression includes contributions from processes beyond second-order perturbation theory; the order of such processes is encoded in the \emph{size} of the matrix. In the following sections we provide further details on how the effective Hamiltonian was obtained in the three cases studied in the main text.

\section{USC effect I: Details for two photons excited by a single atom}
\begin{figure}[t]
\centering
\includegraphics[width=0.99\linewidth]{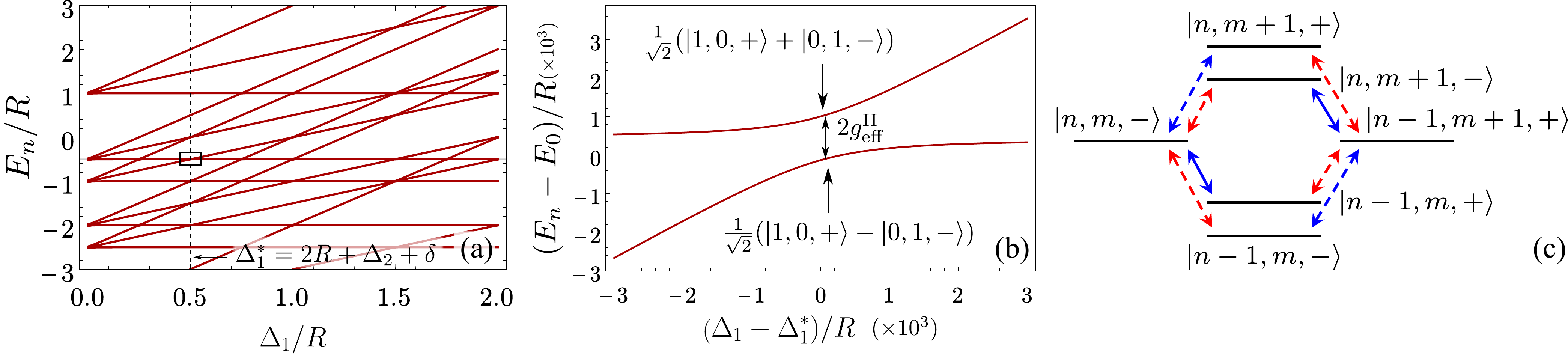}
\caption{USC effect II: Energy-level diagrams and transitions for the process of frequency conversion.
(a) Energy levels $E_n$ for the Hamiltonian in \eqref{eq:H-rotated} as a function of $\Delta_1$. Parameters: $\Omega = 40g$, $\Delta_\sigma = \Delta_\sigma^*$ [see \eqref{eq:optimal-det-freq}],  $\Delta_2 = 2R(f-1)$, with $f=1/4$ and $R=\sqrt{\Omega^2+\Delta_\sigma^2/4}$. The Hilbert space is truncated at 1 photon for simplicity.
(b) Zoom-in on the anti-crossing between the energy levels corresponding to $\ket{1,0,+}$ and $\ket{0,1,-}$. The size of the level splitting at the resonance $\Delta_1 = \Delta_1^* = 2R+\Delta_2+\delta=2Rf+\delta$---where both states have the same energy  in the absence of coupling $E_0 =\Delta_2+R= R(2f-1)$---indicates the strength $g_\mathrm{eff}^\mathrm{II}$ of the effective interaction between these two states.
(c) The transitions in the second-order process that creates the effective coupling between $\ket{n,m,-}$ and $\ket{n-1,m+1,+}$. Blue solid arrows: transitions that conserve the number of excitations. Red dashed arrows: transitions that change the number of excitations by one. Blue dashed arrows: transitions that change the number of excitations by two.
\label{fig:EnergyLevelsAndTransitions_FrequencyConversion}}
\end{figure}

\begin{figure}
\centering
\includegraphics[width=0.99\linewidth]{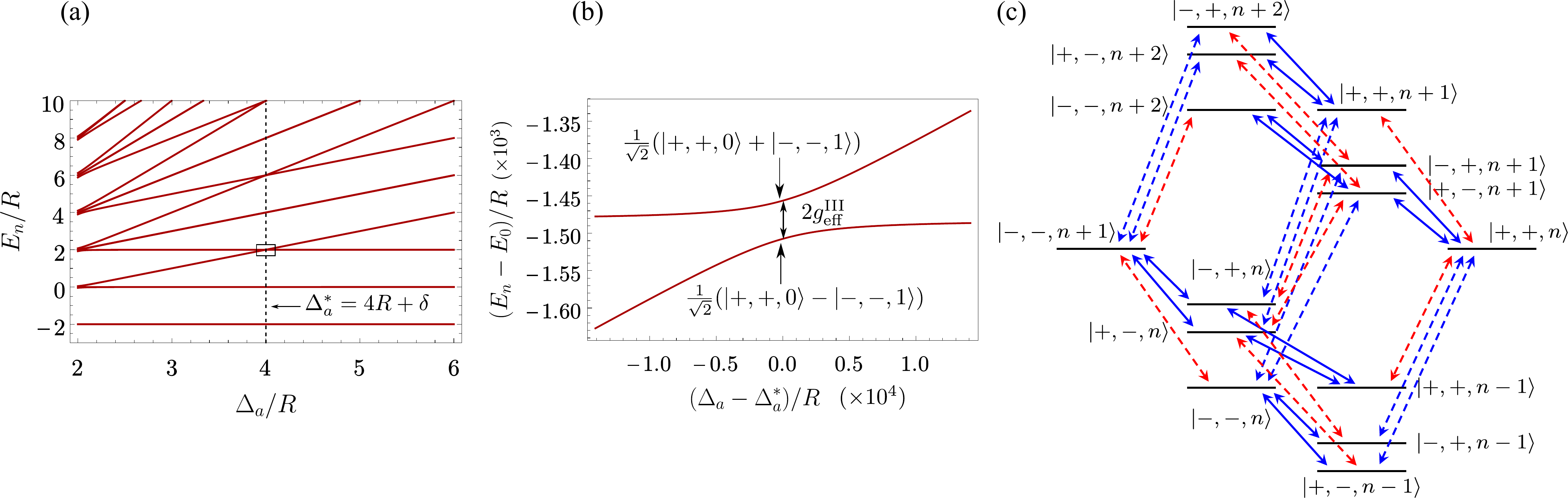}
\caption{USC effect III: Energy-level diagrams and transitions for the process of exciting two atoms with a single photon.
(a) Energy levels $E_n$ for the Hamiltonian in \eqref{eq:initial-H} as a function of $\Delta_a$. Parameters: $\Omega = 20g$, $\Delta_\sigma = \Delta_\sigma^*$ [see \eqref{eq:delta_sigma_opt_III}], with $R=\sqrt{\Omega^2+\Delta_\sigma^2/4}$. The Hilbert space is truncated at 4 photons for simplicity.
(b) Zoom-in on the anti-crossing between the energy levels corresponding to $\ket{+,+,0}$ and $\ket{-,-,1}$. The size of the level splitting at the resonance $\Delta_a = \Delta_a^* = 4R+\delta$---where both states have the same energy  in the absence of coupling $E_0 =2R$---indicates the strength $g_\mathrm{eff}^\mathrm{III}$ of the effective interaction between these two states.
(c) The transitions in the third-order process that creates the effective coupling between $\ket{-,-,n+1}$ and $\ket{+,+,n-1}$. Blue solid arrows: transitions that conserve the number of excitations. Red dashed arrows: transitions that change the number of excitations by one. Blue dashed arrows: transitions that change the number of excitations by two.
\label{fig:EnergyLevelsAndTransitions_OnePhotonTwoAtoms}}
\end{figure}

We consider the following two subspaces, with $N_\mathcal{A}=2$ and $N_\mathcal{B}=12$: 
 \begin{itemize}
 \item  $\mathcal A = \{|n,m,+\rangle, |n+1,m+1,-\rangle \}$,
 \item $\mathcal B =\{ |n+1,m,\pm\rangle, |n,m+1,\pm\rangle,|n+2,m+1,\pm\rangle, |n+1,m+2,\pm\rangle, |n,m-1,\pm\rangle, |n-1,m,\pm\rangle \}  $.\end{itemize}
Note that states such as $|n+2,m+1,\pm\rangle$ do not contribute to the effective coupling between the two states in $\mathcal{A}$, but to the Lamb shifts and dispersive cavity-qubit couplings, through processes such as $|n+1,m+1,-\rangle \rightarrow |n+2,m+1,\pm\rangle \rightarrow |n+1,m+1,-\rangle$.  For simplicity, those processes are neither depicted in Fig.~2 of the main text nor in Fig.~\ref{fig:EnergyLevelsAndTransitions_FrequencyConversion} and \ref{fig:EnergyLevelsAndTransitions_OnePhotonTwoAtoms}.
Using $E=\Delta_1 n + \Delta_2 m + R$, the correction to the effective Hamiltonian in the subspace $\mathcal A$, given by \eqref{eq:effective-H},  has the form:
\begin{equation}
\delta h^\mathrm{I} = \begin{pmatrix}
\chi_1 n + \chi_2 m + \lambda + C & g^\mathrm{I}_\mathrm{eff}\sqrt{(n+1)(m+1)}\\
g^\mathrm{I}_\mathrm{eff}\sqrt{(n+1)(m+1)} & -\chi_1(n+1)-\chi_2(m+1) -\lambda + C
\end{pmatrix},
\end{equation}
which allows us to extract $\chi_1$, $\chi_2$, $\lambda$ and $g^\mathrm{I}_\mathrm{eff}$, and, omitting any overall shift $C$, to write the effective Hamiltonian $H^\mathrm{I}_\mathrm{eff}$ in the general form
\begin{equation}
 H_\mathrm{eff}^\mathrm{I} = \Delta_1 a_1^\dag a_1 + \Delta_2 a_2^\dag a_2 + \mleft( R + \lambda \mright) \tilde \sigma_z \nn + \mleft( \chi_1 a^\dag_1 a_1 + \chi_2 a^\dag_2 a_2 \mright) \tilde \sigma_z + g_\mathrm{eff}^\mathrm{I} \mleft( a_1^\dag a_2^\dag \tilde \sigma + \mathrm{h.c.} \mright) \,. \quad
\end{equation}
The expressions for $g_\mathrm{eff}^\mathrm{I}$, $\chi_1$, $\chi_2$ and $\lambda$  are provided in the main text. 

\section{USC effect II: Details for frequency conversion}

We consider the following two subspaces, with $N_\mathcal{A}=2$ and $N_\mathcal{B}=12$: 
 \begin{itemize}
 \item$\mathcal A = \{|n+1,m,-\rangle,|n,m+1,+\rangle \}$,
 \item$\mathcal{B} = \{ |n+1,m+1,\pm \rangle, |n,m,\pm\rangle, |n+2,m,\pm\rangle, |n+1,m-1\pm\rangle, |n,m+2,\pm\rangle, |\pm,n-1,m\rangle  \}$
 \end{itemize}
Here, the correction that we obtain is 
\begin{equation}
\delta h^\mathrm{II} = \begin{pmatrix}
-(n+1)\chi_1 -m \chi_2 - \lambda + C   &  g_\mathrm{eff}^\mathrm{II}\sqrt{(n+1)(m+1)} \\ 
  g_\mathrm{eff}^\mathrm{II}\sqrt{(n+1)(m+1)}    &  n\chi_1+(m+1)\chi_2 + \lambda + C
\end{pmatrix},
\end{equation}
which allows us to write the general form of the effective Hamiltonian,
\begin{equation}
H_\mathrm{eff}^\mathrm{II} = \Delta_1 a_1^\dag a_1 + \Delta_2 a_2^\dag a_2 + (R + \lambda) \tilde \sigma_z \nn\
+ \mleft( \chi_1 a^\dag_1 a_1 + \chi_2 a^\dag_2 a_2 \mright) \tilde \sigma_z + g_\mathrm{eff}^\mathrm{II} \mleft( a_1^\dag a_2 \tilde \sigma + \mathrm{h.c.} \mright) \, .
\end{equation}

The expression for $g_\mathrm{eff}^\mathrm{II}$ is given in the main text. The dispersive coupling rates are in this case given by
\begin{equation}
\chi_i = g^2 \left( \frac{c^4}{2R+\Delta_i}+\frac{s^4}{2R-\Delta_i}  \right),
\end{equation}
and the Lamb shift is
\begin{equation}
\lambda = \frac{g^2}{2}\left[ \frac{c^4 (4R+\Delta_1+\Delta_2)}{(2R+\Delta-1)(2R+\Delta_2)} +\frac{s^4(4R-\Delta_1-\Delta_2)}{(2R-\Delta_1)(2R-\Delta_2)} \right].
\end{equation}
The dispersive couplings and the Lamb shift make the diagonal elements of $\delta h^\mathrm{II}$ unequal. Since both elements need to be equal in order to achieve complete Rabi oscillations, one needs to introduce a small correction the resonance condition $\Delta_1+\Delta_2 = 2R$. Introducing the correction $\delta$ such that $\Delta_1 = 2R f + \delta$ into the final expression of $H_\mathrm{eff}$ (which implies the approximation of ignoring $\delta$ during the derivation of $H_\mathrm{eff}$), and imposing that the diagonal elements are equal, we are left with the expression for the correction to the general resonance condition:
\begin{equation}
\delta = (2n+1)\chi_1 + (2m+1)\chi_2 + 2\lambda.
\end{equation}
This expression in terms of $\chi_i$ and $\lambda$ coincides with the one obtained for the case of two photons excited by a single atom.
Similarly, driving the qubit on resonance does not maximize $g_\mathrm{eff}^\mathrm{II}$ either. Optimizing the angle gives
\begin{equation}
\theta^* (f \lessgtr \frac{1}{2} ) = \arccos \mleft[ \frac{\sqrt{3 + 2 f \pm \sqrt{9 - 4f + 4f^2}}}{2 \sqrt{2}} \mright].
\end{equation}
The factor gained with respect to $\theta = \pi / 4$ is also $f$-dependent and has the following expression:
\begin{equation}
\frac{g^\mathrm{II}_\mathrm{eff}(\theta^*)}{g^\mathrm{II}_\mathrm{eff}(\theta=\pi/4)}=-\frac{\left(-6 f+f'+3\right) \sqrt{-2 f \left(2 f+f'-2\right)+f'+3}}{8 \sqrt{2} (2 f-1)}
\end{equation}
where $f'=\sqrt{4 (f-1) f+9}$. For the particular case $f = 1 / 4$, we find 
\begin{equation}
g_\mathrm{eff}^\mathrm{II}(\theta^*) \approx 1.76 \, g_\mathrm{eff}^\mathrm{II} (\theta = \pi / 4).
\end{equation}
Alternatively, we can compute the optimal detuning $\Delta_\sigma$ for a fixed $\Omega$ (instead of fixed $R$), which is experimentally more meaningful given that varying $\Delta_\sigma$ for a fixed $\Omega$ is more straightforward than varying $\theta$ for a fixed $R$:
\begin{equation}
\Delta_\sigma^* = - \Omega \frac{(1 - 2 f)}{\abs{1 - 2 f}} \sqrt{- 2 + \frac{1 - \sqrt{1 - f (1 - f)}\abs{1-2f}}{f (1 - f)}}.
\label{eq:optimal-det-freq}
\end{equation}
For the particular case $f = 1 / 4$, $\Delta^*_\sigma \approx -0.96 \, \Omega$, which leads to
\begin{equation}
g^\mathrm{II}_\mathrm{eff} (\Delta_\sigma^*) \approx 1.52 \, g_\mathrm{eff}^\mathrm{II} (\Delta_\sigma = 0).
\end{equation}

\section{USC effect III: Details for two atoms excited by a single photon}
We consider the following two subspaces, with $N_\mathcal{A}=2$ and $N_\mathcal{B}=12$: 
 \begin{itemize}
 \item $\mathcal A =\{|+,+,n\rangle,|-,-,n+1\rangle \}$,
 \item $\mathcal{B} = \{|+,-,n+1\rangle,|-,+,n+1\rangle,|+,+,n+1\rangle,|+,-,n\rangle,|-,+,n\rangle,|-,-,n\rangle,|+,-,n+2\rangle,|-,+,n+2\rangle,|-,-,n+2\rangle,|-,+,n-1\rangle,|+,-,n-1\rangle,|+,+,n-1\rangle \}$.
\end{itemize} 

Here, the correction that we obtain is 
\begin{equation}
\delta h^\mathrm{III} = \begin{pmatrix}
2n\chi + 2\lambda \ + C   &  g_\mathrm{eff}^\mathrm{III}\sqrt{n+1} \\ 
g_\mathrm{eff}^\mathrm{III}\sqrt{n+1}    &  -2(n+1)\chi - 2\lambda+C
\end{pmatrix}.
\end{equation}

The dispersive coupling and the Lamb shift are given by second-order processes:
\begin{eqnarray}
\chi &=& g^2 \frac{(2 R - \Delta_a) c^4 + (2 R + \Delta_a) s^4}{4 R^2 - \Delta_a^2}, \\
\lambda &=& \chi / 2.
\end{eqnarray}
Setting $\Delta_a = 4 R + \delta$, the optimum resonance condition is given by
\be
\delta = (4 n + 2) \chi + 4 \lambda = 4 (n + 1) \chi.
\ee
For a fixed $R$, we can see that the optimal angle $\theta$ maximizing $g_\mathrm{eff}$ is $\theta^* = \arctan \mleft(\sqrt{\frac{1 + \sqrt{5}}{3 - \sqrt{5}}} \mright) \approx 0.356 \pi $, giving the following maximum value of $g_\mathrm{eff}^{\mathrm{III}}$:
\be
g_\mathrm{eff}^\mathrm{III}(\theta^*) = \sqrt{\frac{11 + 5 \sqrt{5}}{8}} \frac{g^3}{6 R^2} \approx 1.67 g_\mathrm{eff}^\mathrm{III} (\theta = \pi / 4).
\label{eq:geff-max}
\ee
That is, when choosing the optimal angle $\theta^*$ we obtain $1.67 \times$ enhancement with respect to the resonant case $\Delta_\sigma=0$, which corresponds to $\theta = \pi / 4$. 
In a similar way, we can express this in terms of the optimal detuning:
\be
\Delta_\sigma^* = \Omega \sqrt{\frac{14 \sqrt{109} - 122}{45}} \approx 0.73 \Omega.
\label{eq:delta_sigma_opt_III}
\ee
The corresponding value of $g_\mathrm{eff}$ is then given by
\be
g_\mathrm{eff}^\mathrm{III} (\Delta_\sigma^*) \approx -1.3 \frac{g^3}{6 \Omega^2} = 1.3 g_\mathrm{eff}^{\mathrm{III}} (\Delta_\sigma = 0)
\ee
giving $ 1.3 \times$ enhancement with respect to the resonant case, for the same driving amplitude $\Omega$.


\section{Validity of the perturbation theory}
\begin{figure}[t!]
\centering
\includegraphics[width=0.8\linewidth]{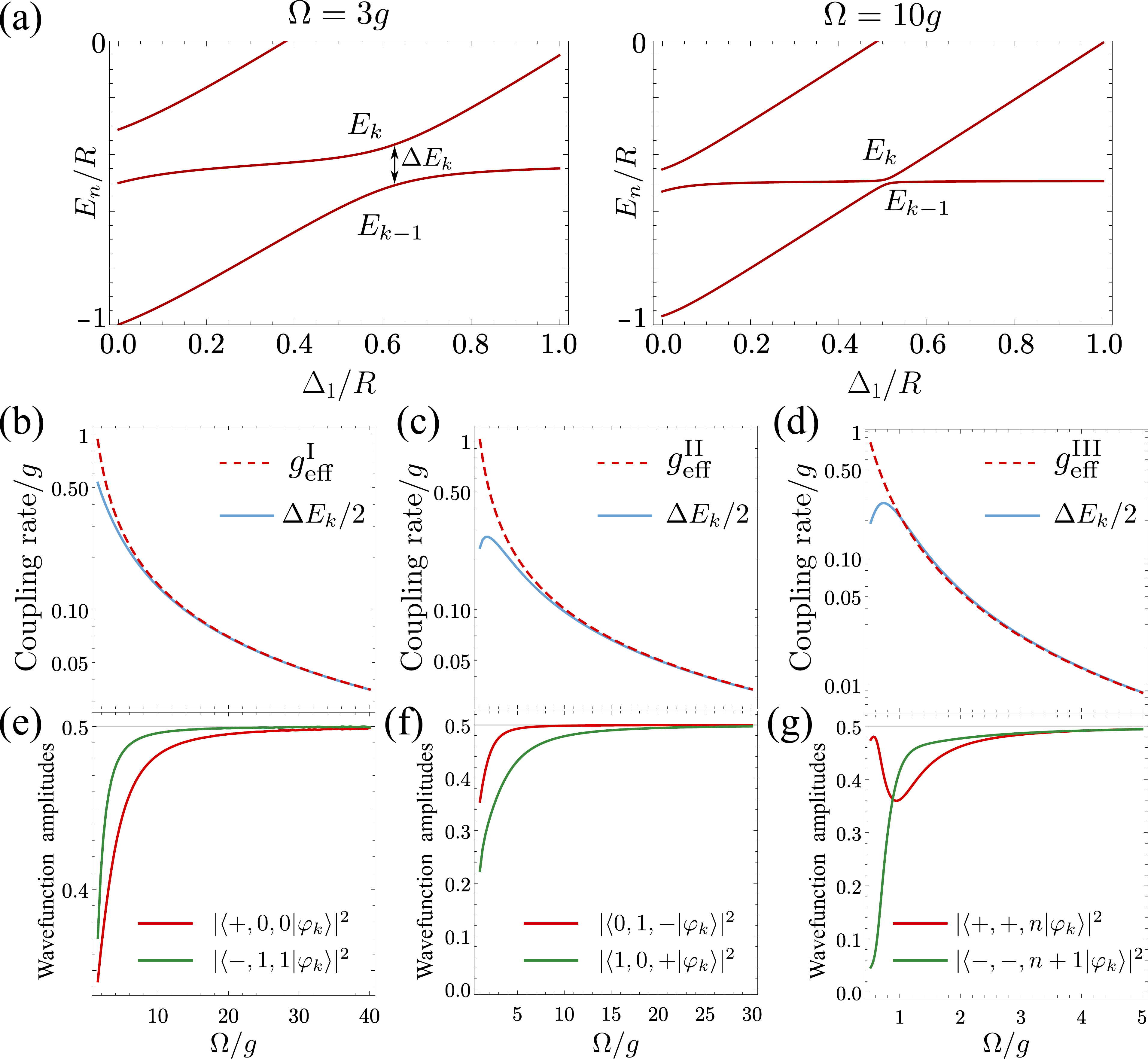}
\caption{Validity of the applied perturbation theory for the three examples. (a) Examples of the avoided crossings at two different values of the perturbation parameter $g/\Omega$. Smaller values of $\Omega$ imply larger splitting. (b-c) Splitting at the avoided crossing (solid) versus $\Omega/g$, compared to the effective rates computed from perturbation theory (dashed). The perturbation theory works when the two curves overlap. (e-f) The overlap between the two eigenstates at the avoided crossing and the two states involved in the nonlinear process; when the perturbation theory starts failing, the overlap is reduced, meaning that eigenstates contain contributions from other states.
\label{fig:PerturbationCheck}}
\end{figure}
%
In this section, we address the question of the validity of the perturbation theory for the three studied USC effects for large values of the perturbation parameter $g/\Omega$. To do so, we study the energy-level splitting $\Delta E_k$ between the two eigenstates $|\varphi_k\rangle$ and $|\varphi_{k-1}\rangle$ at the avoided-level crossing that we associate to each nonlinear process, see Fig.~\ref{fig:PerturbationCheck}~(a). The resulting splitting is compared to the  effective coupling rates $g_\mathrm{eff}^\mathrm{I}$, $g_\mathrm{eff}^\mathrm{II}$ and $g_\mathrm{eff}^\mathrm{III}$ that we have computed from perturbation theory, as we show in Fig.~\ref{fig:PerturbationCheck}~(b-c). In addition, we compute the overlap between the two eigenstates $|\varphi_{k/k-1}\rangle$ and the two states between which we expect the Rabi oscillations to occur in each of the three cases considered in the text, see Fig.~\ref{fig:PerturbationCheck}~(e-g). As $\Omega/g$ is reduced, the effective coupling rates increase. The perturbation theory starts failing at $\Omega/g \lesssim 10$ for cases I and II, and $\Omega/g \lesssim 2$ for case III (which is a third-order process). Below these values of $\Omega/g$, $|\varphi_{k,k-1}\rangle$ stop being composed exclusively of the two isolated states that constitute the desired nonlinear process, and the effective coupling rate predicted from perturbation theory departs from the real half-splittings between these eigenstates.


\section{Effect of decoherence}

\begin{figure}
\centering
\includegraphics[width=0.9\linewidth]{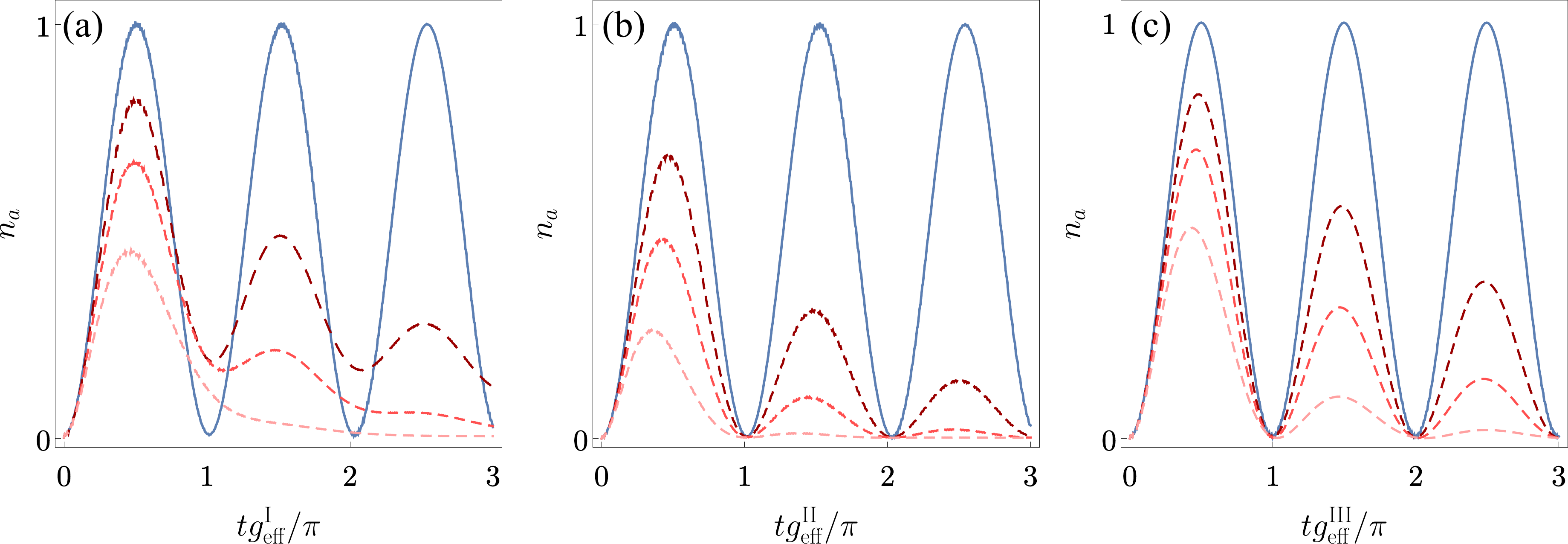}
\caption{Effect of decoherence induced by radiative decay in the cavity modes for the three nonlinear processes considered in this work. In each case, we monitor the population $n_a=\langle a^\dagger a \rangle$ of a cavity undergoing Rabi oscillations under the nonlinear processes with and without decay. Lossless dynamics are represented by straight, blue durves. Dissipative dynamics are shown in dashed curves, with lighter color representing higher decay rates. Decay rates used are $g_\mathrm{eff} \times (0.25,0.5,1)$.
$\Omega/g= 20$ (a), $20$ (b) and $10$ (c).
\label{fig:DampedRabi}}
\end{figure}

Here we provide further study of the effect of decoherence (beyond \tabref{table:experiments} in the main text) computing the dynamics of the nonlinear processes in the presence of cavity losses. This is done by using the standard Lindblad master equation for the dynamics of the density matrix:
\begin{equation}
\dot \rho = -i[H,\rho] + \frac{\gamma_a}{2}\sum_{i=1}^{N_\mathrm{cav}}\left(2 a_i\rho a_i - a_i^\dagger a_i \rho - \rho a_i^\dagger a_i \right).
\end{equation}
where $\gamma_a$ is a cavity decay rate, and the sum runs over the total number of cavities (two in cases I and II, one in case III). In Fig.~\ref{fig:DampedRabi} we show that the effect of decoherence is the expected damping of the Rabi oscillations. The apparent higher robustness to losses of case III (two atoms excited by a single photon) can be explained by the fact that there is only one cavity, which is the only lossy subsystem, as compared to two cavities in cases I and II. We have assumed the cavity decay to be the main decoherence mechanism, therefore ignoring the decay or dephasing of the atoms to simplify the discussion.

\end{document}